\documentclass[sigconf]{acmart}
\renewcommand\footnotetextcopyrightpermission[1]{} 
\usepackage{fancyhdr}
\AtBeginDocument{%
  \providecommand\BibTeX{{%
    \normalfont B\kern-0.5em{\scshape i\kern-0.25em b}\kern-0.8em\TeX}}}

\newtheorem{myDef}{Definition}
\usepackage{subfigure}

\begin{document}

\title{Drug Package Recommendation via
Interaction-aware \\ Graph Induction}

\author{Zhi Zheng$^{1}$, Chao Wang$^{1}$, Tong Xu$^1$, Dazhong Shen$^{1}$, Penggang Qin$^1$,\\Baoxing Huai$^{2}$, Tongzhu Liu$^{3}$, Enhong Chen$^{1}$}
\affiliation{%
\institution{$^1$Anhui Province Key Lab of Big Data Analysis and Application, University of Science and Technology of China\\
            $^2$Huawei Technologies, 
       $^3$The First Affiliated Hospital of USTC}}
\affiliation{\{zhengzhi97, wdyx2012, sdz, qinpg\}@mail.ustc.edu.cn, \{tongxu, cheneh\}@ustc.edu.cn,\\huaibaoxing@huawei.com, liutongzhu2015@126.com}

\titlenote{Tong Xu is the corresponding author.}

\renewcommand{\shortauthors}{Zhi Zheng, et al.}
\renewcommand{\shorttitle}{Drug Package Recommendation via Interaction-aware Graph Induction}




\begin{abstract}
Recent years have witnessed the rapid accumulation of massive electronic medical 
records (EMRs), which highly support the intelligent medical services such as drug 
recommendation. However, prior arts mainly follow the traditional recommendation 
strategies like collaborative filtering, which usually treat individual drugs as 
mutually independent, while the latent interactions among drugs, e.g., synergistic 
or antagonistic effect, have been largely ignored. To that end, in this paper, 
we target at developing a new paradigm for drug package recommendation with 
considering the interaction effect within drugs, in which the interaction effects 
could be affected by patient conditions. Specifically, we first design a 
pre-training method based on neural collaborative filtering to get the initial 
embedding of patients and drugs. Then, the drug interaction graph will be initialized 
based on medical records and domain knowledge. Along this line, we propose a new 
Drug Package Recommendation (DPR) framework with two variants, respectively DPR on 
Weighted Graph (DPR-WG) and DPR on Attributed Graph (DPR-AG) to solve the problem, 
in which each the interactions will be described as signed weights or 
attribute vectors. In detail, a mask layer is utilized to capture the impact of 
patient condition, and graph neural networks (GNNs) are leveraged for the final 
graph induction task to embed the package. Extensive experiments on a real-world 
data set from a first-rate hospital demonstrate the effectiveness of our DPR 
framework compared with several competitive baseline methods, and further support 
the heuristic study for the drug package generation task with adequate performance.
\end{abstract}

\ccsdesc[500]{Information systems~Data mining}

\keywords{Drug Recommendation, Package Recommendation, Graph Neural Network}

\maketitle

\section{INTRODUCTION}
With the growth of population and the intensification of population aging, 
people's demand for high-quality medical services continues to rise, 
and the pressure on the medical workers is increasing. Moreover, 
certain public health emergencies such as the outbreak of COVID-19, 
will also have a huge impact on the medical system. Meanwhile, 
artificial intelligence (AI) technologies have shown enormous potential to reduce 
human labor. Therefore, if AI technologies could be effectively utilized to realize 
intelligent diagnosis and drug recommendation clinically, 
it will greatly improve 
the overall quality of medical services.


Fortunately, with the popularization of information technology in the medical industry, 
electronic medical records (EMRs) have been widely used in major hospitals,  
which powerfully support the downstream intelligent applications like 
medical image analysis \cite{litjens2017survey,dilsizian2014artificial}, 
chronic disease management \cite{garcia2007new, martin2008smart}, medical  
text analysis \cite{afzal2017mining, miller2017using}, etc. 
However, due to the limitation of data and technology, 
drug recommendation based on EMR is 
still largely unexplored. In terms of data, similar to traditional recommendation system, 
drug recommendation is sensitive to data quality, but it is hard to get reliable medical 
data sources. Moreover, most patients have only been recorded once or 
several times in EMR database, 
which makes it hard to utilize conventional personalized recommendation methods 
based on user preference analysis. In terms of 
technology, it is very important for the recommender system to 
consider both drug effect and the interaction between drugs at the same time, 
and give the patient a suitable drug package, which contains multiple drugs. However, 
most of existing studies generally rely on traditional methods such as 
collaborative filtering \cite{zhang2015cadre} to solve this problem. 
Due to the lack of item relation data for interaction analysis, there are limits for 
these methods to achieve satisfactory performance in 
practical applications. 

\begin{figure}[t]
    \centering
    \includegraphics[width=0.45\textwidth]{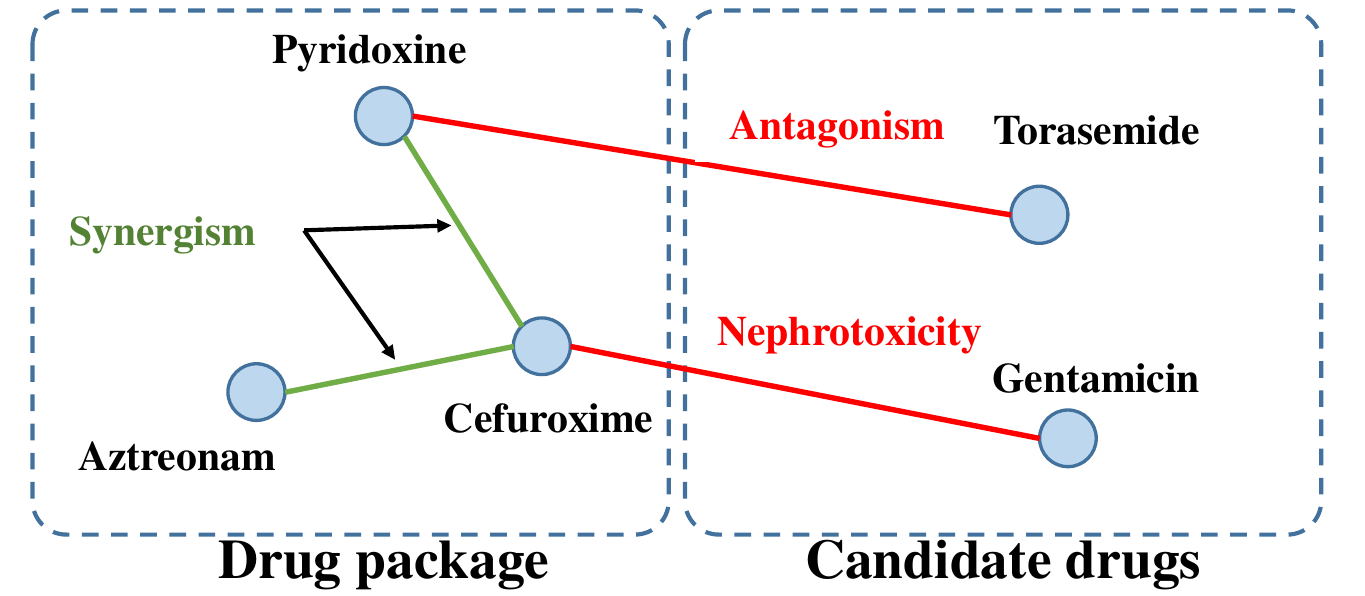}
    \caption{An example for a patient with kidney disease.}
    \label{fig:intro}
    \end{figure}

In order to address the above challenges, in this paper, we aim to develop a new paradigm for drug 
package recommendation with the awareness of drug interaction. The rationale behind this is that 
the interaction between drugs will influence the effect of the drug package, and
the impact of drug interaction on drug effect will be further 
affected by patient conditions. We illustrate this by a patient with kidney disease 
as shown in Figure~\ref{fig:intro}. The drug package for this patient contains three 
drugs, respectively pyridoxine, aztreonam and cefuroxime. Cefuroxime is synergistic with 
the other two drugs, which can improve the effect of the drug package. Torasemide is 
antagonistic with pyridoxine, so it is not included in the package. Furthermore, 
the combination of cefuroxime and 
gentamicin has a synergistic antibacterial effect, but at the same time it may increase 
nephrotoxicity, so it is not suitable for this patient.

Along this line, we first design a pre-training model to get the embedding of patients 
and drugs based on neural collaborative filtering (NCF). 
Then we collect drug interaction data from public online dataset 
and divide drug pairs into three categories with the help of domain experts, respectively No Interaction, Synergism and Antagonism.  
After that, we propose to represent drug packages as graphs 
based on the labeled data. 
Furthermore, we propose 
a Drug Package Recommendation (DPR) framework with two variants. The first one, 
namely DPR on Weight Graph (DPR-WG), regards the effect of 
drug interaction as graph edge weights, while the second one, 
DPR on Attributed Graph (DPR-AG), utilizes edge attribute vectors to describe the 
influence of drug interaction. In both two models, we exploit a mask layer to capture 
the impact of the patient condition on the drug package representation, and 
Graph Neural Networks (GNNs) are leveraged for the final graph induction task to embed the package. 
Finally, extensive experiments on a real-world dataset from a first-rate hospital 
demonstrate the effectiveness of our DPR framework compared with several 
competitive baseline methods, and further support the heuristic study for the 
drug package generation task with adequate performance.

Specifically, the major contributions of this paper can be summarized as follows:

\begin{itemize}
    \item We develop a new paradigm to represent drug packages as graphs based on 
    drug interaction classification. 
    \item We design a drug package recommendation framework with two variants, 
    which can integrate drug interaction information based on graph induction.
    \item We propose to utilize a mask layer to capture 
    the impact of patient condition on the drug package representation.
    \item We conduct extensive experiments on a real-world data set from a first-rate hospital, 
    which clearly validate the effectiveness of our DPR framework and reveal some interesting rules 
    based on the derived insights on patient conditions and drug interaction.
\end{itemize}

\section{RELATED WORK}
In this section, we will summarize the related works as following three categories, 
respectively drug recommendation system, package recommendation system, and graph 
neural networks.

\subsection{Drug Recommendation System} Recommendation systems have been widely used in 
a variety of applications like social networking and e-commerce. The methods can be 
broadly classified into two categories, respectively neighborhood-based collaborative 
filtering methods based on similar users or items \cite{adomavicius2005toward}, 
and model-based methods, particularly latent factor models that factorize the 
user-item matrix into user factors and item factors \cite{koren2009matrix}. 
Recent recommender systems have been further advanced by the significant contribution 
from deep learning \cite{zhang2019deep, he2017neural, xue2017deep}, where user 
preferences and item characteristics can be learned in deep architectures. 
Based on these technologies, some methods focusing on drug recommendation have been put forward. 
For example, 
\cite{zheng2020multi} introduces a LDA-based contextual collaborative model called Medicine-LDA 
to integrate the multi-source information. 
\cite{zhang2014towards} constructs a heterogeneous graph which includes patients and 
drugs, and describes a novel recommendation system based on label propagation. \cite{chiang2018drug} 
develops a joint model with a recommendation component and an ADR label prediction component  
to recommend a set of to-avoid drugs. 
With the increasing emergence of knowledge graph, some researchers have extracted 
information from medical database like \cite{law2014drugbank} to build up giant medical 
knowledge graphs. Based on these knowledge graphs, \cite{wang2017safe} proposes to 
jointly embed diseases, drugs and patients into a shared lower dimensional space, and 
decomposes the drug recommendation into a link prediction process. 
However, these models lack the ability to recommend drugs as a package, and the studies 
on drug interaction are not thorough enough.

\subsection{Package Recommendation System} Most recommendation research concentrates 
on recommending one item to users at a time. However, in many real world scenarios, the 
platform needs to show users a set of items, in other words, a package (or a bundle). 
Several efforts have been made to solve this problem. Some studies turn 
this problem into optimization problems like 0-1 Knapsack problem, and provide 
some approximate solutions due to the NP-Hardness 
\cite{parameswaran2009recommendations, lappas2009finding, deng2013complexity, zhu2014bundle}. 
\cite{liu2011personalized} puts forward a Tourist-Area-Season topic model and proposes 
a cocktail approach on personalized travel package recommendation. 
\cite{bai2019personalized} proposes a bundle generation network which decomposes the 
problem by derterminantal point processes. \cite{pathak2017generating} develops a 
model which utilizes the trained features of an item recommendation model to learn the 
personalized ranking over bundles. \cite{chen2019matching} contributes a neural network 
solution based on factorized attention network to aggregate the item embeddings in a 
package. \cite{chang2020bundle} proposes a model based on graph neural network which 
explicitly models the interaction and affiliation between users, bundles, and items by 
unifying them into a heterogeneous graph. However, these models neglect the 
different types of interactions between items, which prevents them from capturing 
satisfactory performance for drug package recommendation.

\subsection{Graph Neural Networks} Recently, many studies on extending deep learning 
approaches for graph data have emerged. Unlike standard neural networks, GNNs 
retain a state that can represent information from its neighborhood with 
arbitrary depth. For example, \cite{kipf2016semi} presents graph convolutional network (GCN) 
for semi-supervised learning on graph data via an approximation of 
spectral graph convolutions. \cite{hamilton2017inductive} presents GraphSAGE to 
generate node embeddings by sampling and aggregating features from the local 
neighborhoods of nodes. \cite{velivckovic2017graph} presents graph attention networks 
(GATs) which leverage masked self-attentional layers to address the shortcomings 
of methods based on graph convolutions. 
\cite{10.5555/3305381.3305512} further presents that the 
essence of existing GNNs is to learn a message passing algorithm and an aggregation 
procedure to compute a function of the entire input graph, and 
reformulates existing models into a single common framework called Message Passing 
Neural Networks (MPNNs). 
With the strong power of learning structure, GNNs have been widely applied 
in many fields. For example, \cite{zhang2020large, li2020competitive} 
utilize graph data and graph neural networks for competitive analysis. 
\cite{liu2020exploiting} propose a deep model to integrate structural and temporal 
social contexts to address the dynamic social-aware recommendation task. 

\section{PRELIMINARIES}
In this section, we first introduce the real-world dataset used in our study, and then 
propose the problem formulation of drug package recommendation.

\subsection{Data Description and Preprocessing}
\label{sec:data}
\begin{figure}[tb]
  \centering
  \includegraphics[width=\linewidth]{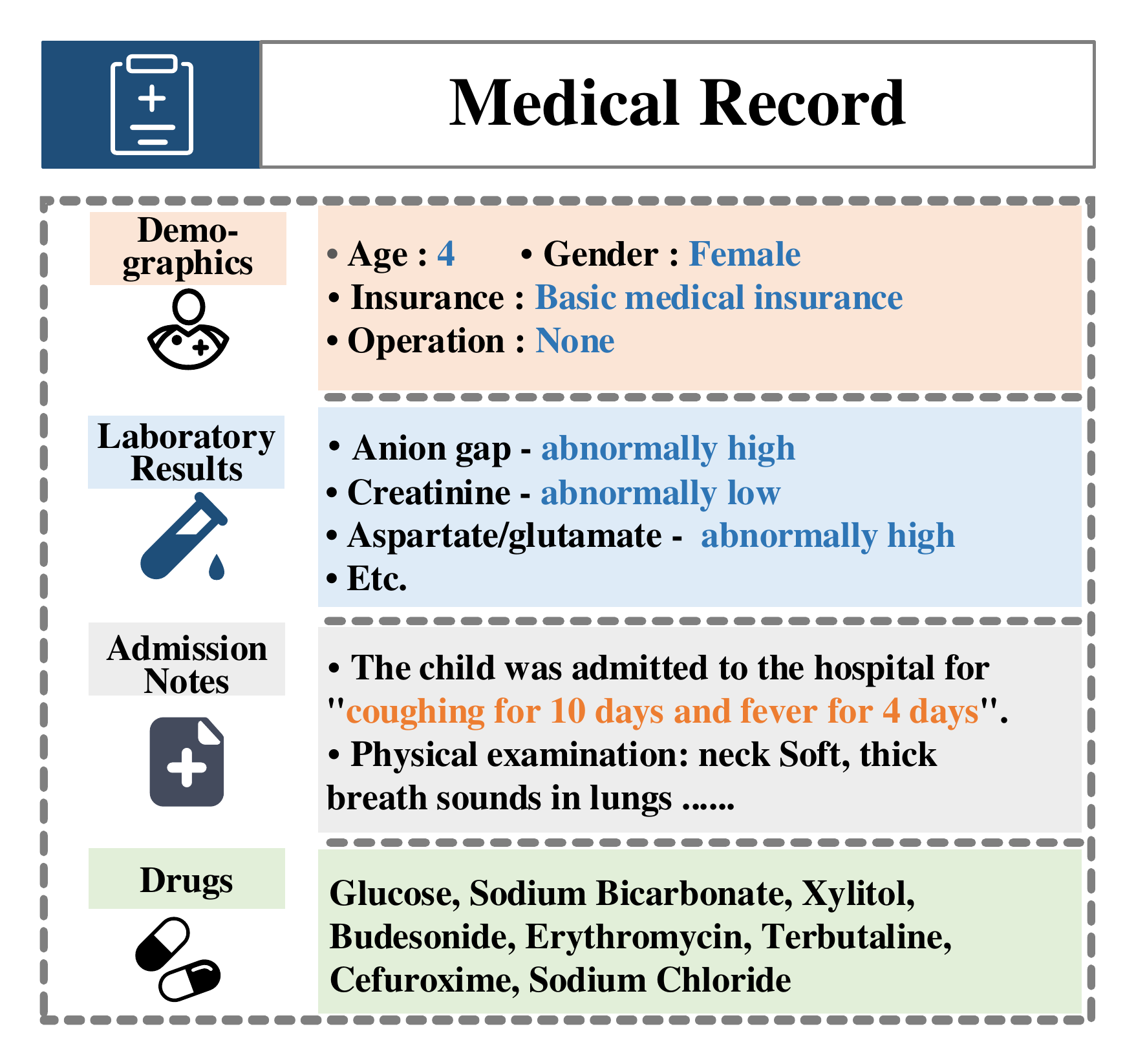}
  \caption{An example of the medical record in our dataset.}
  \label{fig:record}
\end{figure}

The EMR dataset used in this paper comes from the electronic medical record database 
of a first-rate hospital in China. 
As shown in Figure~\ref{fig:record}, each medical record contains the following information:
\begin{itemize}
    \item \textbf{Demographics}. Demographics are formatted data including basic patient information, 
    such as patient's gender, age, type of 
    medical insurance, whether surgery has been performed, etc. This information 
    provides guidance for doctors to prescribe, for example, some drugs are not 
    suitable for children, while some drugs are only covered by certain medical insurance, etc.  
    \item \textbf{Laboratory results}. A laboratory test is a procedure in which the hospital 
    takes a sample of the patient's body fluid or body tissue to get information of the 
    patient's health. The laboratory results are shown as the patient's values and normal 
    values for laboratory items. For example, "glucose value: 77 mg/dL, normal value: 65-99 mg/dL".
    \item \textbf{Admission notes}. An admission note is part of a medical record that documents 
    the patient's status including physical examination findings, reasons why the patient 
    is being admitted for inpatient care to a hospital, and the initial instructions 
    for the patient's care.
    \item \textbf{Drugs}. This information includes all of the drugs used during the patient's hospital stay.
\end{itemize}

In order to integrate and utilize the above multi-source heterogeneous data, we conduct 
the following preprocessing steps. First, for the demographics, we convert them 
into documents, e.g., "Gender : Male, Age : Teenager". Second, 
for the laboratory results, we divide 
the results into three levels, respectively normal, abnormally high and abnormally 
low according to the given 
normal values. We then extract all abnormal test results (abnormally high and low) 
from the results and converted them 
into documents, e.g., "glucose value : abnormally high, lipid panel : abnormally high". 
After that, we merge the demographic documents and laboratory result documents, 
namely disease documents. Finally, 
for the admission notes, we remove all the punctuation and meaningless characters, 
and adjust all of the admission notes in the dataset to the same length by 
padding and cut-off.

For the purpose of studying the interaction between drugs, we collect data from 
two large online pharmaceutical 
knowledge bases, i.e., 
DrugBank\footnote{https://go.drugbank.com/releases/latest} and YaoZhi\footnote{https://db.yaozh.com/interaction}, 
where users can check drug properties and drug-drug interaction. 
The drug interaction information in these two databases are stored in text 
format based on some certain 
templates. We further classify the templates into three categories with the help of domain experts, 
respectively No Interaction, Synergism and Antagonism. No Interaction means there is no interaction 
between two drugs. Synergism means the combination of two drugs 
can lead to enhanced drug effect, and Antagonism is the opposite. Table~\ref{tab:Interaction} 
shows some examples of different drug interactions. Note that the interaction can be 
directed, for example, if drug A can increase the effect of drug B, then the direction is from 
A to B. Moreover, for most of the drug pairs, we cannot confirm whether 
there is any type of interactions between them, so we leave them as unlabeled. 
Section~\ref{sec:construction} will further discuss how to exploit these labeled and 
unlabeled data.

Finally, we pick out the EMR records containing more than one drug and 
we get totally 158,556 EMR records with complete information. More detailed 
statistics of our data are shown in Table~\ref{tab:Statistics}.
\begin{table}[tb]
  \caption{Statistics of our dataset.}
  \label{tab:Statistics}    
  \begin{tabular}{c|c}
  \hline
  Discription & Number \\\hline
  The number of records & 158,556 \\
  The number of drugs & 1,428 \\
  The number of words in disease document  & 1,242 \\
  The average size of drug packages  & 18 \\
  The number of aligned drugs& 565 \\
  The number of drug pairs with No Interaction  & 2,560 \\
  The number of drug pairs with Synergism  & 22,986 \\
  The number of drug pairs with Antagonism  & 6,389 \\
  \hline
  \end{tabular}
  
\end{table}

\begin{table*}[tb]
\caption{Examples of drug interaction labeling.}
\label{tab:Interaction}
\begin{tabular}{ccccc}
\toprule
Drug A&Drug B&Description&Classification&Direction\\
\midrule
Amoxicillin & Oseltamivir  & No Interaction & No Interaction & Bidirection\\
Dipyridamole & Valsartan & Dipyridamole may increase the antihypertensive activities of Valsartan. & Synergism & A to B\\
Repaglinide & Doxepin & Doxepin may decrease the hypoglycemic activities of Repaglinide. & Antagonism & B to A\\
\bottomrule
\end{tabular}
\end{table*}

\subsection{Problem Formulation}
Based on the above EMR and drug interaction data, here we introduce the problem 
formulation of drug package recommendation. For facilitating illustration, Table~\ref{tab:notations} 
lists some important mathematical notations used throughout this paper.

Suppose there are $N$ patients and $M$ drugs in the training set. 
Based on the above preprocessing method, for patient $i$, 
we can construct the disease document and turn it into one-hot encoding form as 
$\mathcal{W}_i=\left\{w_{i,1},w_{i,2},\dots,w_{i,p}\right\}$, 
where $w_{i,\cdot}$ is the 0/1 indicator value for a 
demographic feature or a lab result. In addition, we can formulate the admission note as 
$\mathcal{T}_i=\left\{t_{i,1},t_{i,2},\dots,t_{i,q}\right\}$, 
where $t_{i,\cdot}$ is a word in the processed 
admission note. In this way, the patient $i$ can be 
expressed as a patient description $\mathcal{U}_i=\left\{\mathcal{W}_i,\mathcal{T}_i\right\}$. 
We also have the drug package $\mathcal{P}_i=\left\{d_{i,1},d_{i_2},\dots,d_{i,s}\right\}$, 
where $d_{i,\cdot}$ is a drug that patient $i$ used. 
Moreover, based on the labeled drug 
interaction data, we can construct the drug relation matrix 
$\mathcal{R}\in \mathbb{R}^{M \times M}$, where $\mathcal{R}_{ij}$ represents the interaction 
between $d_i$ and $d_j$, namely 
\textbf{0 for No Interaction, 1 for Synergism, 2 for Antagonism and -1 for unknown.} 
Note that the direction is from $d_i$ to $d_j$. 
Along this line, the problem of drug package recommendation can 
be formulated as:

\begin{myDef}[Drug Package Recommendation]
Given a set of patient descriptions $\{\mathcal{U}_1,\mathcal{U}_2,\dots,\mathcal{U}_N\}$ 
with the corresponding drug packages $\{\mathcal{P}_1,\mathcal{P}_2,\dots,\mathcal{P}_N\}$, 
and the drug relation matrix $\mathcal{R}$, 
the goal of drug package recommendation is to get a personalized scoring function for each 
patient: $f_u:\mathcal{P} \to \mathbb{R}$.
\end{myDef}

Note that the cold start patients and packages are very common in our 
drug package recommendation problem. 
For example, a new patient comes to the hospital or a doctor prescribes a new drug package. 
This requires the model to score a package based on the patient condition and 
the effect of drug packages, making the problem radically different 
from traditional recommendation based on user-item interaction matrix.

\begin{table}[tb]
  \caption{Mathematical notations.}
  \label{tab:notations}
  \resizebox{\columnwidth}{!}{
  \begin{tabular}{c|l}
  \hline\hline
  Symbol & Description                                        \\ \hline \hline
  $N,M$ & The number of patients and the number of drugs; \\
  $\mathcal{P}_i$ & The drug package of patient $i$; \\
  $\mathcal{W}_i$ & The disease document of patient $i$; \\
  $\mathcal{T}_i$ & The admission note of patient $i$; \\
  $\mathcal{U}_i$ & The patient discription of patient $i$; \\
  $\mathcal{G}_i$ & The drug package graph of patient $i$; \\
  $\mathcal{R}$ & The drug relation matrix; \\
  $\Theta$ & Model Parameters;\\
  $d_{j}$ & The $j$th drug in the entire drug set; \\
  $d_{i,\cdot}$ & Drug in the drug package of patient $i$; \\
  $w_{i,\cdot}$ & Indicator value in the disease document of patient $i$; \\
  $t_{i,\cdot}$ & Word in the admission note of patient $i$; \\
  $MLP\left( \cdot \right)$ & Multilayer Perceptron with ReLU Activation Function.\\
  \hline \hline
  \end{tabular}
  }
\end{table}

\section{TECHNICAL DETAILS}

\begin{figure*}[tb]
    \centering
    \includegraphics[width=\linewidth]{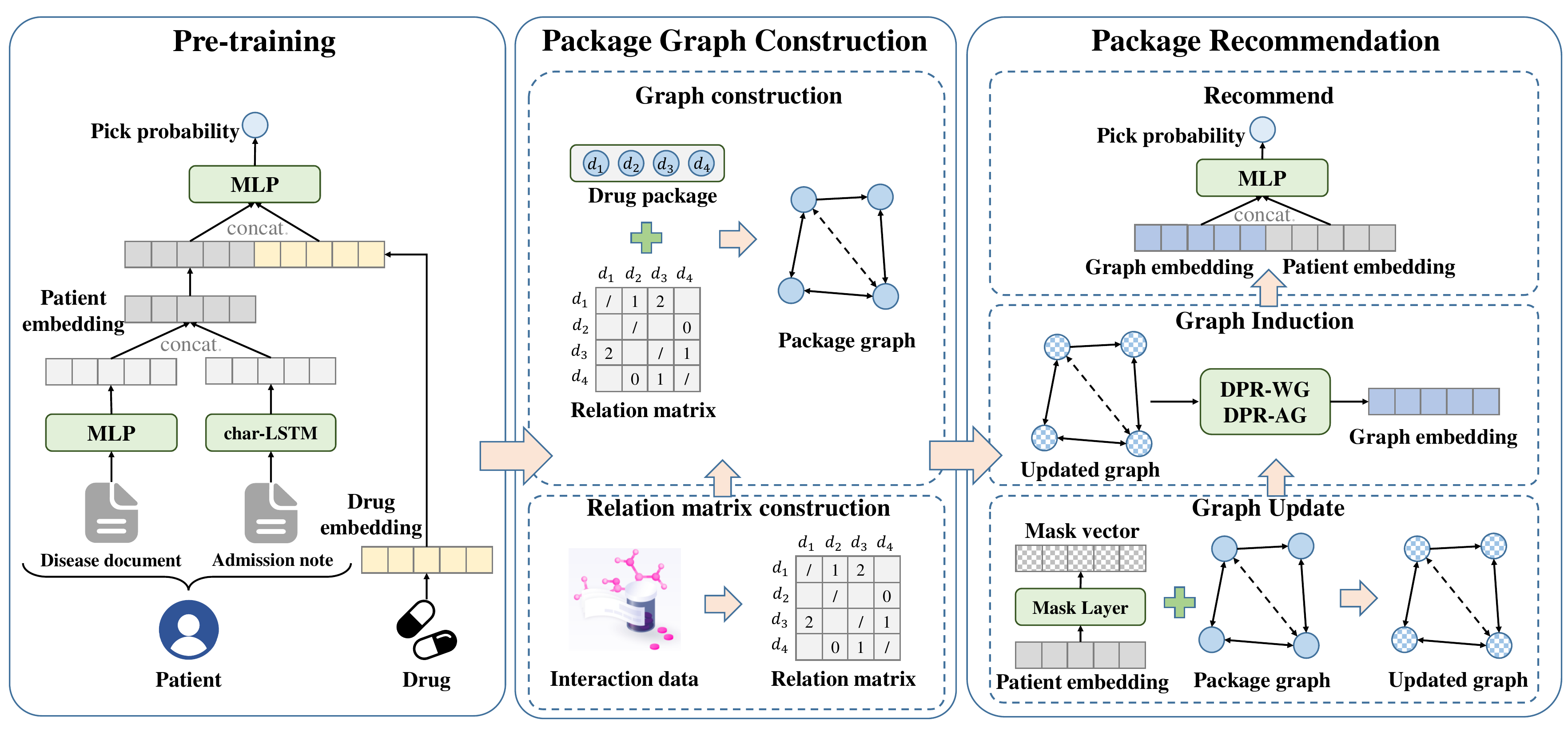}
    \caption{A framework overview of the drug package recommendation system.}
    \label{fig:model}
  \end{figure*}

In this section, we will introduce the framework of our model in detail. As shown in 
Figure~\ref{fig:model}, our framework mainly consists of three components, i.e., 
pre-training, package graph construction, and drug 
package recommendation. Specifically, we first design a pre-training 
method based on neural collaborative filtering to get the initial embedding of 
patients and drugs. Then, we propose to construct drug package graphs based on the 
medical records and domain knowledge. Finally, a novel Drug Package Recommendation (DPR) 
framework with two variants are proposed to solve the drug package recommendation problem.

\subsection{Pre-training}
\label{sec:pre-training}
A patient's description consists of two heterogeneous parts, and a drug package consists 
of several drugs. 
In order to recommend drug packages,  
we first need to get the embeddings of drugs and patients. Therefore, we propose a 
pre-training method as follows.

First, we propose a hybrid method  
to get the patient embedding $\mathbf{u}$ based on 
patient description $\mathcal{U}=\{\mathcal{W},\mathcal{T}\}$, which 
can be split into two steps. To be specific, in the first step,   
we extract the feature of the patient's disease document by MLP as:
\begin{equation}
\mathbf{m}_{w} = MLP\left(\mathcal{W}\right).
\end{equation}

In the second step, we associate each word $t_k$ in patients' admission notes with a word 
embedding vector $\mathbf{x}_k$. By this way we can convert $\mathcal{T}$ to a sequence of 
vectors $(\mathbf{x}_{1},\mathbf{x}_{2},\dots,\mathbf{x}_{q})$. 
Then we input the sequence into char-LSTM \cite{lample-etal-2016-neural} as:
\begin{equation}\begin{array}{l}
\mathbf{i}_{t}=\sigma\left(\mathbf{W}_{x i} \mathbf{x}_{t}+\mathbf{W}_{h i} \mathbf{h}_{t-1}+\mathbf{W}_{c i} \mathbf{c}_{t-1}+\mathbf{b}_{i}\right), \\
\mathbf{f}_{t}=\sigma\left(\mathbf{W}_{x f} \mathbf{x}_{t}+\mathbf{W}_{h f} \mathbf{h}_{t-1}+\mathbf{W}_{c f} \mathbf{c}_{t-1}+\mathbf{b}_{f}\right), \\
\mathbf{c}_{t}=\mathbf{f}_{t} \odot \mathbf{c}_{t-1}+\mathbf{i}_{t} \odot \tanh \left(\mathbf{W}_{x c} \mathbf{x}_{t}+\mathbf{W}_{h c} \mathbf{h}_{t-1}+\mathbf{b}_{c}\right), \\
\mathbf{o}_{t}=\sigma\left(\mathbf{W}_{x o} \mathbf{x}_{t}+\mathbf{W}_{h o} \mathbf{h}_{t-1}+\mathbf{W}_{c o} \mathbf{c}_{t}+\mathbf{b}_{o}\right), \\
\mathbf{h}_{t}=\mathbf{o}_{t} \odot \tanh \left(\mathbf{c}_{t}\right).
\end{array}\end{equation}

We get the final time step output $\mathbf{h}_{q}$ as the embedding of $\mathcal{T}$, 
and the patient embedding $\mathbf{u}$ is the concatenation of the two parts:
\begin{equation}
\mathbf{u} = \left[\mathbf{m}_{w}||\mathbf{h}_{q}\right].
\end{equation}

Second, we associate each drug $d_j$ with a randomly initialized embedding $\mathbf{d}_j$ which 
directly projects drug one-hot ID to the latent space. Finally, 
We utilize Neural Collaborative Filtering (NCF) framework \cite{he2017neural} 
and Bayesian Personalized Ranking (BPR) loss \cite{10.5555/1795114.1795167} 
to train the above embeddings and models. Specifically, 
for patient $i$, we get a patient-drug predictive model by 
feeding patient embedding $\mathbf{u}_i$ and drug embedding $\mathbf{d}_j$ into a 
matching model:
\begin{equation}
\hat{r}_{ij} = MLP\left(\left[\mathbf{u}_{i}||\mathbf{d}_{j}\right] \right),
\end{equation} 
Then we adopt BPR loss as:
\begin{equation}
L=\sum_{i=1}^{N}\sum_{j \in \mathcal{P}_i}\sum_{l \notin \mathcal{P}_i} -\ln \sigma\left(\hat{r}_{ij}-\hat{r}_{il}\right) + \lambda \left\|\Theta \right\|_{2}^{2},
\end{equation} 
where $d_j$ is in drug package and $d_l$ is not. We minimize the loss function forcing the prediction 
$\hat{r}_{ij}$ to be larger than $\hat{r}_{il}$. $\sigma(\cdot)$ is the sigmoid function,  
and $\Theta$ is the parameter set. $L_{2}$ regularization is applied to prevent overfitting.

\subsection{Package Graph with Message Passing}
\label{sec:package_graph}
Compared with traditional item recommendation, the core problem of drug package 
recommendation is how to get the representation of drug packages 
considering the interaction between drugs. 
Therefore, in this section, we propose to utilize graph models to solve this problem. 
To be specific, we first present a method to convert the drug packages into package 
graphs. Then, we formulate the message passing framework which will be further 
utilized for the graph induction task.


\subsubsection{Package Graph Construction.}
\label{sec:construction}
For drug package $\mathcal{P}$, 
we define a corresponding package graph 
$\mathcal{G}=\{\mathcal{V}, \mathcal{E}\}$, 
where $\mathcal{V}$ is the node set and $\mathcal{E}$ is the edge set. 
Each specific node $v \in \mathcal{V}$ is associated with corresponding 
drug embedding $\mathbf{d}$. Each directed edge $e_{vu} \in \mathcal{E}$ also has its 
attribute, and its form will change with different methods, which will be discussed in 
later sections.

The topology structure of the package graph $\mathcal{G}$, i.e., whether edge 
$e_{vu}$ should exist, needs to be defined. Theoretically, since any pair of drugs may 
have drug interaction, the package graph $\mathcal{G}$ 
should be a complete graph, where all nodes are connected with each other. 
However, this will make the time complexity of graph induction 
increases from $O\left(n\right)$ 
to $O\left(n^2\right)$ owing to the pairwise interaction. Furthermore, 
we find that the frequency of drug co-occurrence obeys a long-tailed distribution, 
which means most of the drug pairs have no clear relationship. Therefore,  
we propose the following two criterions to define the topology of a package graph. 
For nodes $v, u$: 
$1)$ If $\mathcal{R}_{vu} \ne -2$, which means this drug paired has been labeled 
in Section~\ref{sec:data}, then edge $e_{vu}$ exists. $2)$ Calculate the 
co-occurrence proportion $p_{ij} = num_{ij} / num_i$, where ${num_i}$ means the number of packages containing drug $i$, 
and $num_{ij}$ means the number of packages containing both drug $i$ and drug $j$. If 
$p_{ij}$ is bigger than a threshold value, then edge $e_{vu}$ exists.

\subsubsection{Message Passing on Package Graph.}
We propose to exploit the MPNN \cite{10.5555/3305381.3305512} framework for making 
use of the package graphs constructed in the last section. 
MPNN is a general approach to describe GNNs, which inductively learns a node 
representation by recursively aggregating and transforming the feature vectors of 
its neighboring nodes. A per-layer update of the MPNN model in our setting involves 
message passing, message aggregation, and node representation updating, which can be 
expressed as:
\begin{equation}\mathbf{m}_{v u}^{(l)}=\operatorname{MESSAGE}(\mathbf{h}_{u}^{(l-1)}, \mathbf{h}_{v}^{(l-1)}, \mathbf{e}_{v u}),\end{equation}
\begin{equation}\mathbf{M}_{u}^{(l)}=\operatorname{AGGREGATION}(\{\mathbf{m}_{v u}^{(l)}, \mathbf{e}_{v u}\} \mid v \in \mathcal{N}(u)\}),\end{equation}
\begin{equation}\mathbf{h}_{u}^{(l)}=\operatorname{UPDATE}(M_{u}^{(l)}, \mathbf{h}_{u}^{(l-1)}),\end{equation}
where $\mathbf{m}_{v u}^{(l)}$ is the message vector passing from $v$ to $u$, $\mathbf{h}_{u}^{(l)}$ is the representation of node $u$ on the
layer $l$; $\mathbf{e}_{v u}$ is the attribute corresponding to edge $e_{vu}$. 
$\mathcal{N}(u)$ is the neighborhood of node $u$ from where it collects information to
update its aggregated message $\mathbf{M}_{u}$. $\mathbf{h}_{u}^{(0)}$ is 
initialized by corresponding drug embedding $\mathbf{d_i}$, and we also express it as 
$\mathbf{d_u}$ for facilitating illustration.

\subsection{Drug Package Recommendation}
After the formulation of package graphs and massage passing neural networks, 
we can finish the graph induction task, i.e., get the embedding of the package graph, 
based on the MPNN framework and further solve the drug package recommendation problem. 
The key to obtain effective representation of the drug package graph is 
to utilize the edge attributes to capture the interaction between the drugs. Therefore, 
we propose the following two ways to formulate the edge attributes in package graphs from 
two different point of views. First, since the two major interactions in our dataset, 
respectively Synergism and Antagonism, are opposite to each other, we can simply exploit 
signed edge weights to describe the drug interaction intensity. Second, if we 
expect our model to be more generic, we can define the edge attributes as vectors 
which contain the information about the type of interaction. Along this line, 
we propose our Drug Package Recommendation (DPR) model with two variants, 
respectively DPR on Weighted Graph (DPR-WG) and 
DPR on Attributed Graph (DPR-AG) in the following sections.

\subsubsection{DPR on Weighted Graph}
\label{sec:DPR-WG}
\begin{figure}[t]
    \centering
    \includegraphics[width=0.45\textwidth]{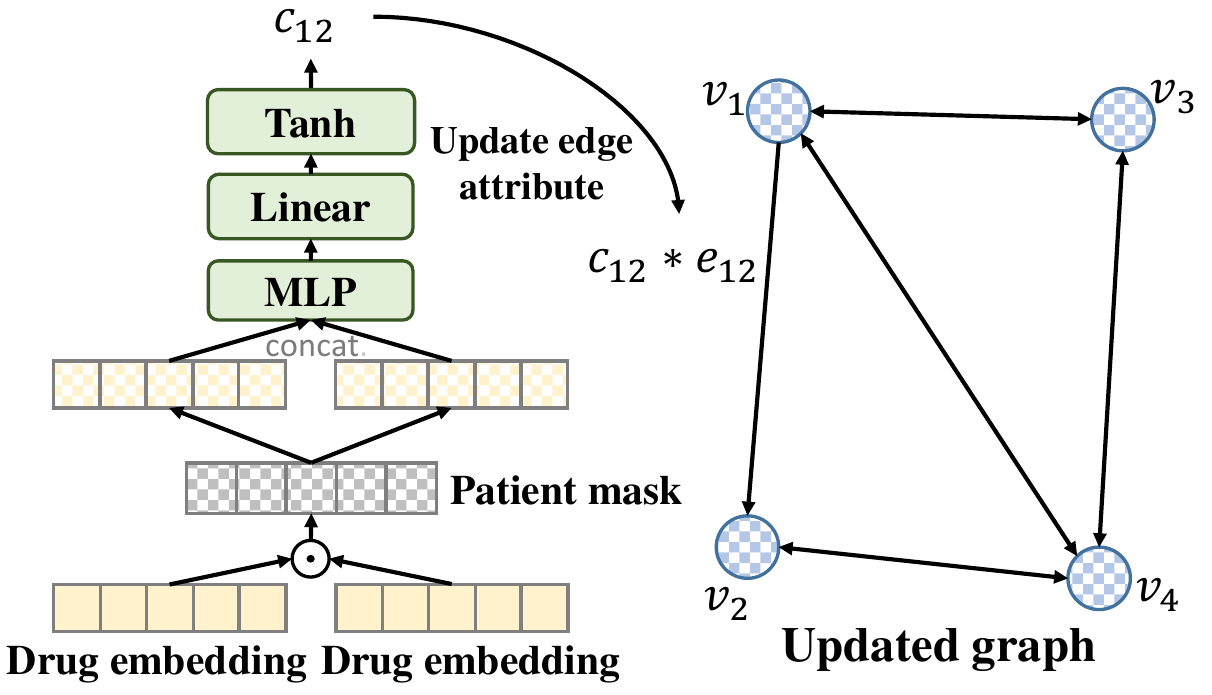}
    \caption{Edge attribute updating progress of DPR-WG.}
    \label{fig:DPR-WG}
    \end{figure}

In DPR-WG, we present to convert a 
package graph $\mathcal{G}$ into a weighted graph by assigning real numbers to edge 
attributes, i.e., $e_{vu} \in \mathbb{R}$. 
Specifically, for edge $e_{vu}$ in a package graph $\mathcal{G}$, we initialize the 
edge attribute as: 

$$\mathbf{e}_{vu}=\left\{
\begin{array}{rcl}
  1   &    &      \mathcal{R}_{vu}=1,\\
  -1   &  &      \mathcal{R}_{vu}=2,\\
  p_{vu}   &  &       otherwise.
\end{array} \right. $$
Note that we set edge attribute 
for $e_{vu}$ even if $\mathcal{R}_{ij}=0$ or $\mathcal{R}_{ij}=-1$, since the 
interaction data may be incomplete or incorrect. 

As previously stated, the impact of drug interaction will also be affected 
by patient condition. 
Inspried by \cite{wang2019mcne}, 
we propose to utilize a mask layer to extract a mask 
vector from a patient's embedding $\mathbf{u}$, and get the conditional drug embedding 
$\hat{\mathbf{d}}_u$ as follows:
\begin{equation}
\hat{\mathbf{d}}_u = \sigma\left(MLP\left(\mathbf{u}\right)\right) \odot \mathbf{d}_u,
\end{equation}
where the mask layer is formed as $\sigma\left(MLP\left( \cdot \right)\right)$, and 
the mask vector $\sigma\left(MLP\left(\mathbf{u}\right)\right)$ 
plays the role of feature selecting on the drug embeddings. 
$\odot$ represents the element-wise product of two vectors. 
Then, for edge $e_{vu}$, we can calculate a contextual impact factor $c_{vu}$ as:
\begin{equation}
c_{vu} = Tanh(a^\top MLP([\hat{\mathbf{d}}_{u}||\hat{\mathbf{d}}_{v}] )),
\end{equation}
where $a^\top$ is a row vector which has the same length with the MLP output.  
The contextual impact factor $c_{vu}$ reflects the impact of the the patient condition 
on the drug interaction between $d_u$ and $d_v$. After the above calculation, we can 
update the edge attribute as 
$\hat{\mathbf{e}}_{vu} = c_{vu} * \mathbf{e}_{vu}$. Figure~\ref{fig:DPR-WG} shows the 
updating progress in detail.

Then, we can form the GNN layer using edge weight for filtering as the following steps:
\begin{equation}\mathbf{m}_{v u}^{(l)}=W_{1}^{(l-1)} \mathbf{h}_{v}^{(l-1)}, \end{equation}
\begin{equation}\mathbf{M}_{u}^{(l)}=\sum_{v \in \mathcal{N}(u)} GRU\left(\hat{\mathbf{e}}_{vu}\mathbf{m}_{v u}^{(l)}, \mathbf{h}_{u}^{(l-1)}\right),\end{equation}
\begin{equation}\mathbf{h}_{u}^{(l)}=MLP\left(W_{0}^{(l-1)} \mathbf{h}_{u}^{(l-1)}+\mathbf{M}_{u}^{(l)}\right),\end{equation}
where $W$ denotes the model’s parameters to be learned, and GRU denotes the 
gated recurrent neural network \cite{chung2014empirical}. 
We set the dimension of all the 
layers equal to the dimension of 0th layer.

Now we can utilize the formed GNN layer for the graph induction task. 
Note that different from general sparse graphs, 
a drug package graph is a graph which is dense enough. 
Therefore, we only need one layer 
of GNN to extract almost all the information we expected, and there is no need for 
high-order neighbors, which will be discussed later. 
For each node $v$, we have the initial node embedding $\mathbf{d}_v$ 
and the corresponding hidden representation $\mathbf{h}_v$ from the GNN layer. 
Following \cite{li2015gated}, the package 
graph embedding can be formed as:
\begin{equation}\label{induction}\mathbf{g}=\sum_{v \in V} \sigma\left(MLP\left(\left[\mathbf{d}_v||\mathbf{h}_v\right]\right)\right) \odot\left(MLP\left(\left[\mathbf{d}_v||\mathbf{h}_v\right]\right)\right).\end{equation}
Again, we utilize NCF framework and BPR loss to train the model. For patient $i$, we have 
the patient embedding $\textbf{u}_i$ and the corresponding package graph embedding $\mathbf{g}_i$. The loss 
function can be formed as follows, where the MLP model is the final prediction model:
\begin{equation}
\begin{aligned}
L=\sum_{i=1}^{N}\sum_{j \ne i}\ -\ln \sigma\left(MLP\left(\left[\mathbf{u}_i||\mathbf{g}_i\right]\right)-MLP\left(\left[\mathbf{u}_i||\mathbf{g}_j\right]\right)\right) + \lambda \left\|\Theta \right\|_{2}^{2},
\end{aligned}
\end{equation}

\subsubsection{Drug Package Recommendation on Attributed Graph}
\label{sec:DPR-AG}
\begin{figure}[t]
  \centering
  \includegraphics[width=0.45\textwidth]{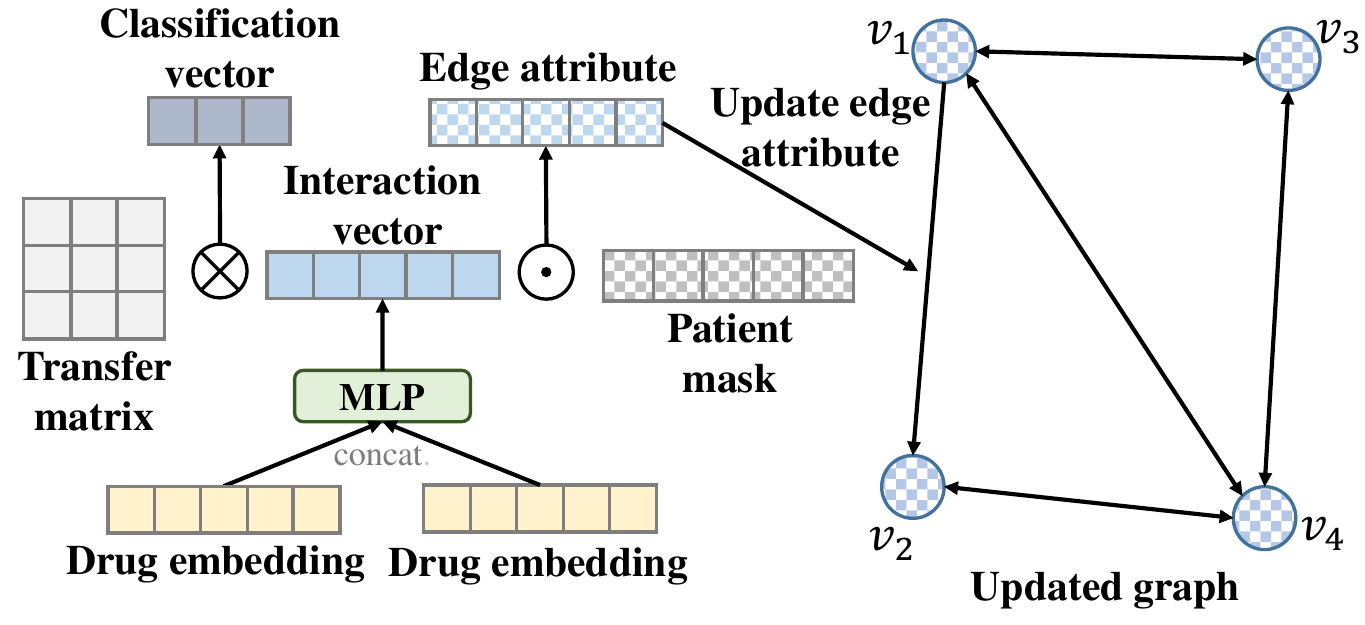}
  \caption{Edge attribute updating progress of DPR-AG.}
  \label{fig:DPR-AG}
  \end{figure}
In DPR-AG, the package graph $\mathcal{G}$ is formed as an attributed graph, 
where both nodes and edges have corresponding attribute vectors. 

Specifically, for edge $e_{vu}$ and the corresponding drug embedding 
$\mathbf{d}_v, \mathbf{d}_u$, we first form the edge attribute vector $\mathbf{e}_{vu}$ 
as the interaction vector between $\mathbf{d}_v$ and $\mathbf{d}_u$, 
which is calculated by MLP model as:
\begin{equation}
\mathbf{e}_{vu}=MLP\left(\left[\mathbf{d}_v||\mathbf{d}_u\right]\right).
\end{equation}
Then, we utilize the mask layer again to update the edge attributes by 
adding the impact of patient’s condition on the interaction vector as follows:
\begin{equation}
\hat{\mathbf{e}}_{vu}=\sigma\left(MLP\left(\mathbf{u}\right)\right) \odot \mathbf{e}_{vu}.
\end{equation}
The detailed updating progress is shown in Figure~\ref{fig:DPR-AG}. 
Based on the above steps, we can form the GNN layer as 
the following steps, note that the settings are the same as DPR-WG in the former section:
\begin{equation}\mathbf{m}_{v u}^{(l)}=W_{1}^{(l-1)} \hat{\mathbf{e}}_{vu}^{(l-1)}, \end{equation}
\begin{equation}\mathbf{M}_{u}^{(l)}=\sum_{v \in \mathcal{N}(u)} \mathbf{m}_{v u}^{(l)},\end{equation}
\begin{equation}\mathbf{h}_{u}^{(l)}=MLP\left(W_{0}^{(l-1)} \mathbf{h}_{u}^{(l-1)}+\mathbf{M}_{u}^{(l)}\right).\end{equation}
After getting the package graph embedding by equation~\ref{induction}, we can form the 
loss function for DPR-AG. The essential difference between DPR-WG and DPR-AG is that, 
in DPR-WG, the prior knowledge is leveraged explicitly by initializing the edge weights 
according to the relation matrix $\mathcal{R}$. On the contrary, we propose to utilize the 
prior knowledge implicitly in DPR-AG. Specifically, we design a hybrid loss function as:
\begin{equation}
\begin{aligned}
L&=\sum_{i=1}^{N}\sum_{j \ne i}\ -\ln \sigma\left(MLP\left(\left[\mathbf{u}_i||\mathbf{g}_i\right]\right)-MLP\left(\left[\mathbf{u}_i||\mathbf{g}_j\right]\right)\right) \\ 
 &- \sum_{i=1}^{N}\sum_{\substack{u,v \in \mathcal{G}_i\\\mathcal{R}_{uv} \ne -1}} \ln\left(softmax\left(\mathbf{e}_{vu}^{\top}\mathbf{Q}\right)_{\mathcal{R}_{uv}}\right)  + \lambda \left\|\Theta \right\|_{2}^{2},
\end{aligned}
\end{equation}
where the MLP model is the final prediction model. $\mathbf{Q} \in \mathbb{R}^{D\times 3}$ is the transfer 
matrix to transform the edge attribute $\mathbf{e}_{vu}$ into classification probabilities, 
where $D$ is the dimension of $\mathbf{e}_{vu}$. 
We add cross entropy loss to the loss function, which aims to force the edge attribute 
$\mathbf{e}_{vu}$ to contain the interaction type information.

\section{EXPERIMENTS}
In this section, we evaluate the proposed model with a number of competitive 
baselines. Meanwhile, many discussions and
case studies on drug package recommendation will be presented.

\subsection{Experimental Settings}
We omit the dataset description in this section since it has been introduced in 
Section~\ref{sec:data}. Other experimental settings will be described in the following 
parts. 
\subsubsection{Baselines and Evaluation Metrics.} To evaluate the performance of our 
models for drug package recommendation, we selected 
a number of state-of-art methods as baselines. Specifically, we first chose two popular 
traditional recommendation approaches, and several state-of-art package recommendation 
models as follows:
\begin{itemize}
\item \textbf{NCF}~\cite{he2017neural}: NCF is a state of art deep neural networks on 
recommendation system, which replacing the inner product in  
matrix factorization with a neural architecture. This model recommends top $K$ drugs 
as packages for the patients in test sets based on the patient embeddings, where $K$ is 
the average size of drug packages.
\item \textbf{NN}: This method utilizes the pretrained patient embeddings based on NCF, 
and returns the drug package corresponding to the Nearset Neighbor (NN) by calculating the 
cosine similarity of patient embeddings.
\item \textbf{Package2vec}: \cite{wan2018representing} proposes to utilize 
Item2vec \cite{barkan2016item2vec} for enhancing the item embeddings in a package
, and we extend Item2vec following \cite{le2014distributed} to get the embedding 
of a package. 
NCF framework and BPR loss are utilized to train the package recommendation model.
\item \textbf{LDA}~\cite{blei2003latent}: This method utilizes the LDA model to get the 
embedding of a package and uses the same framework as Package2vec to recommend packages.
\item \textbf{BR}~\cite{pathak2017generating}: BR is a package recommendation method which aggregates 
item latent vectors to get the package embeddings based on package size and item compatibility.
\item \textbf{DAM}~\cite{chen2019matching}: DAM is the state-of-art neural network 
architecture for package recommendation which utilizes factorized attention network 
to get the embedding of packages.
\item \textbf{GNN}: This method is a simplified variant 
of our models, which only uses the package graph structure and ignore the edge attributes. 

\end{itemize}

It is worth noting that the drug package recommendation is much different from general 
recommendation since there is no fixed users in our task. Therefore, in all of the baseline 
methods, we exploited the patient embedding model proposed in Section~\ref{sec:pre-training} 
to get the representation of patients. Another problem is how to generate packages for patients 
in test set since most of the models are discriminant. Therefore, we proposed that 
except for the 
NCF model which can generate packages itself, all the remaining models only pick out the best 
package from a candidate set, and the candidate set consists of drug packages from 
10 most similar patients. The similarity was calculated by the cosine similarity between patient 
embeddings. Evaluation metrics including Precision, 
Recall and F1-score were utilized to compare the performance of the models.

\subsubsection{Implementation Details.} We implemented our model by PyTorch\footnote{https://pytorch.org/}  
and Pytorch Geometric\footnote{https://github.com/rusty1s/pytorch\_geometric}. The parameters 
were all initialized using Kaiming \cite{he2015delving} initialization.
For the pre-training 
model, we set the output dimension of the MLP, the dimension of char embeddings, and the
hidden size of the LSTM as 32, while the dimension of patient embeddings was set as 64.
For the construction of package graph, we set the threshold value of co-occurrence proportion 
as 0.01. For the BPR loss used in this paper, we used negative sampling 
to train the model and set the negative sampling ratio as 10, which 
means 10 negative samples for one positive sample. For all the MLP models used in this paper, 
we set the dimension of hidden layers as 128. In the process of model training, 
we used the Adam optimizer \cite{DBLP:journals/corr/KingmaB14} for 
parameter optimization. We set learning rate as 0.001 and mini-batch 
size as 256. The parameters of baselines were set up similarly as our 
method and were all tuned to be optimal to ensure fair comparisons. For the dataset splitting,   
we divided our dataset into 80\%/10\%/10\% training/validation/test and 
we report performance on the test set for the model that performed best on the 
validation set.
    
\begin{table}[tb]  
  \centering  
  \caption{The performance of each model.}  
  \label{tab:Performance}  
    \begin{center}  
        \begin{tabular}{c|ccc}  
            \hline  
          model & Precision & Recall & F1-score \\\hline
          NCF& 0.3812 & 0.5442 & 0.4200 \\
          \hline
          NN& 0.4890  & 0.4985 & 0.4732 \\
          Package2vec& 0.4846  & 0.5268 & 0.4857 \\
          LDA& 0.5014 & 0.5219 & 0.4904 \\
          BR& 0.5068 & 0.5106 & 0.4879 \\
          DAM& 0.5254& 0.5107 & 0.4979 \\
          GNN&0.5085 &0.5288 &0.5009 \\
          \hline
          DPR-WG& 0.5133 & \textbf{0.5488} & 0.5137 \\
          DPR-AG& \textbf{0.5260} & 0.5407 & \textbf{0.5162} \\
      \hline 
        \end{tabular}  
    \end{center}  
    
  \end{table}
  
\subsection{Discussions}
\label{sec:overall}
\subsubsection{Overall Performance. }To demonstrate the effectiveness of our drug package recommendation framework, we  
compared DPR-WG and DPR-AG with all the baselines, and the results are shown in 
Table~\ref{tab:Performance}. From the results, we can get several observations:

First, the performance of our models surpasses most of the 
baseline methods on different evaluation metrics.
This clearly proves the effectiveness 
of our DPR framework based on package graph construction and message passing neural networks. 
Furthermore, our models obtain much higher 
recall than baselines, which indicates our models are more likely 
to prevent doctors from neglecting certain factors in practical application. 

Second, the performance of NCF model is the worst, since this method based on 
collaborative filtering prefers to recommend items with higher popularity, and cannot 
model the drugs as a whole. This clearly verifies the necessity for the studies of package recommendation systems.

Third, the GNN model which only leverages the graph topological structure to exchange information between 
different drugs cannot achieve comparable result with our model. However, this model 
surpasses all the other baselines. This verifies the effectiveness of constructing 
package graphs to capture the interaction between drugs, 
and futher indicates the effectiveness of our method for the graph induction process.

Last but not least, the results of the models except NCF are close to each other, since 
patients with similar condition are more likely to use similar drugs.

\subsubsection{Ablation Study.}
To further validate the effectiveness of each component of our models, 
we also designed some simplified variants of our models as follows:
\begin{itemize}
\item\textbf{DPR-WG-Context}: This method is a simplified variant of DPR-WG which only utilizes 
the edge attributes initialized by the drug interaction matrix and ignores the influence of 
the patient condition.
\item\textbf{DPR-WG-Type}: This method is a simplified variant of DPR-WG which only uses the 
contextual impact factor as edge attributes and ignores the drug interaction type. 
\item\textbf{DPR-AG-Mask}: This method is a simplified variant of DPR-AG which deletes the 
mask layer in the calculation process.
\item\textbf{DPR-AG-Type}: This method is a simplified variant of DPR-AG which deletes the 
cross entropy loss in the loss function. In this way, the edge attributes dose not contain 
the information of drug interaction type.
\end{itemize}

The results of ablation study are shown in Table~\ref{tab:ablation} 
from which we can draw the following conclusions. 
First, DPR-WG performs better than the two variants. This indicates 
that both the 
contextual impact factors and the initial edge weights are significant, which 
clearly verifies our assumption that patient 
condition will influence the interaction effect between drugs. 
Second, DPR-AG also performs better than the two variants, which verifies that 
both parts of 
drug interaction type and mask vectors are effectual, and the mask layer 
we proposed can effectively extract the feature of patient condition. 

\begin{table}[tb]  
    \centering  
    \caption{The results of ablation study.}  
    \label{tab:ablation}  
      \begin{center}  
          \begin{tabular}{c|ccc}  
              \hline  
              
            model & Precision & Recall & F1-score \\\hline
            DPR-WG-Context&0.5126 & 0.5330&0.5053 \\
            DPR-WG-Type& 0.5126 & 0.5377&0.5074\\
            DPR-AG-Mask& 0.5152 & 0.5342&0.5061 \\
            DPR-AG-Type& 0.5154 & 0.5317 & 0.5056 \\ 
            \hline
            DPR-WG& 0.5133 & 0.5488 & 0.5137 \\
            DPR-AG& 0.5260 & 0.5407 & 0.5162 \\
        \hline 
          \end{tabular}  
      \end{center}  
 
    \end{table}

\begin{figure}[t]
  \centering
  \includegraphics[width=0.4\textwidth]{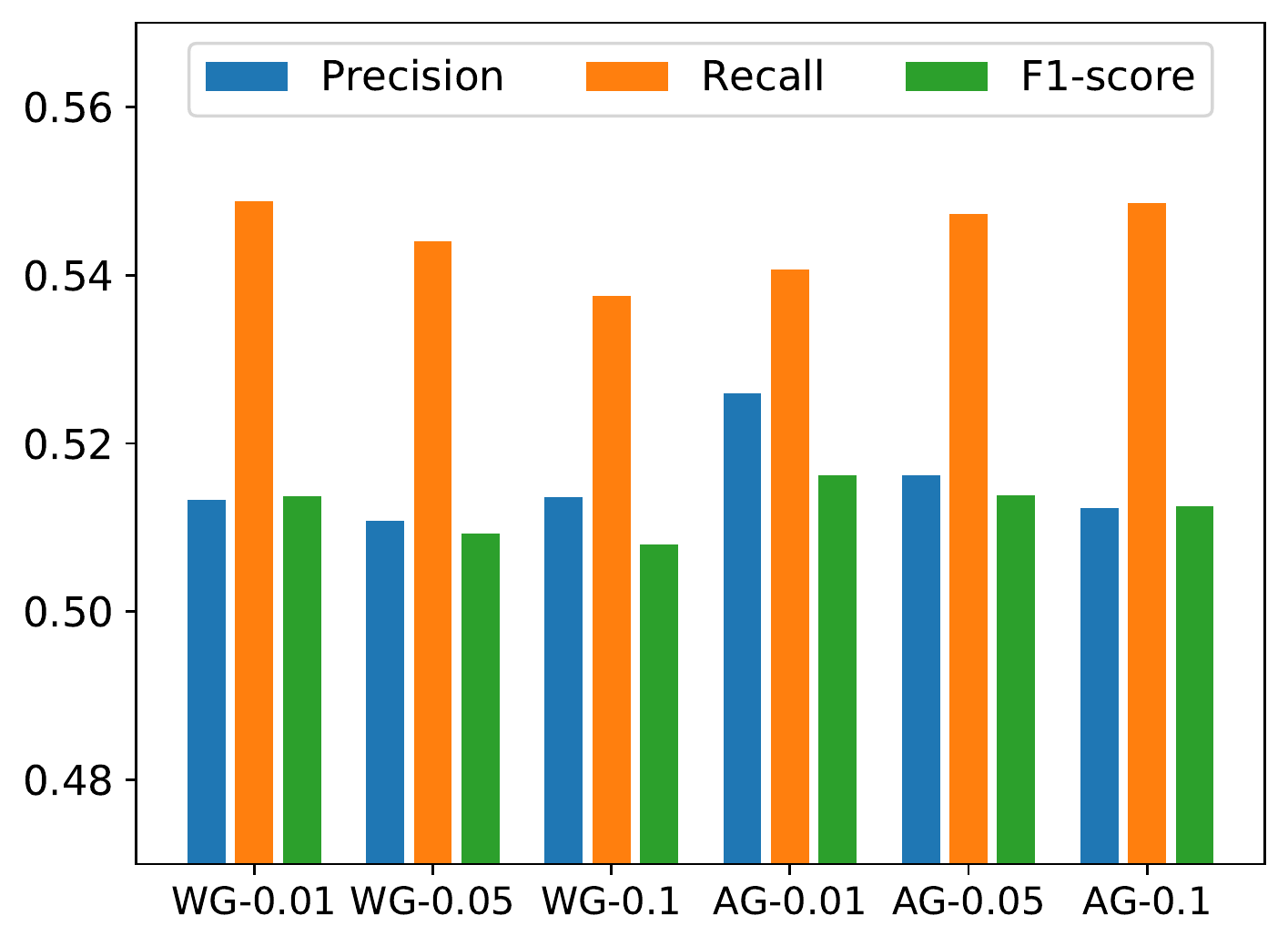}
  \caption{The preformance of our models with different co-occurrence proportion threshold.}
  \label{fig:threshold}
  \end{figure}

\begin{table}[tb]  
      \centering  
      \caption{The preformance of our models with different number of GNN layers.}
      \label{tab:layer}  
        \begin{center}  
            \begin{tabular}{c|ccc}  
                \hline  
              model & Precision & Recall & F1-score \\\hline
              DPR-WG-1& 0.5133 & 0.5488 & 0.5137 \\
              DPR-WG-2& 0.4994 & 0.5582 & 0.5100 \\
              DPR-AG-1& 0.5260 & 0.5407 & 0.5162 \\
              DPR-AG-2& 0.5139 & 0.5457 & 0.5128  \\
          \hline 
            \end{tabular}  
        \end{center}  
   
\end{table}

\begin{figure}
\centering
\subfigure[DPR-WG]{
    \includegraphics[scale=0.25]{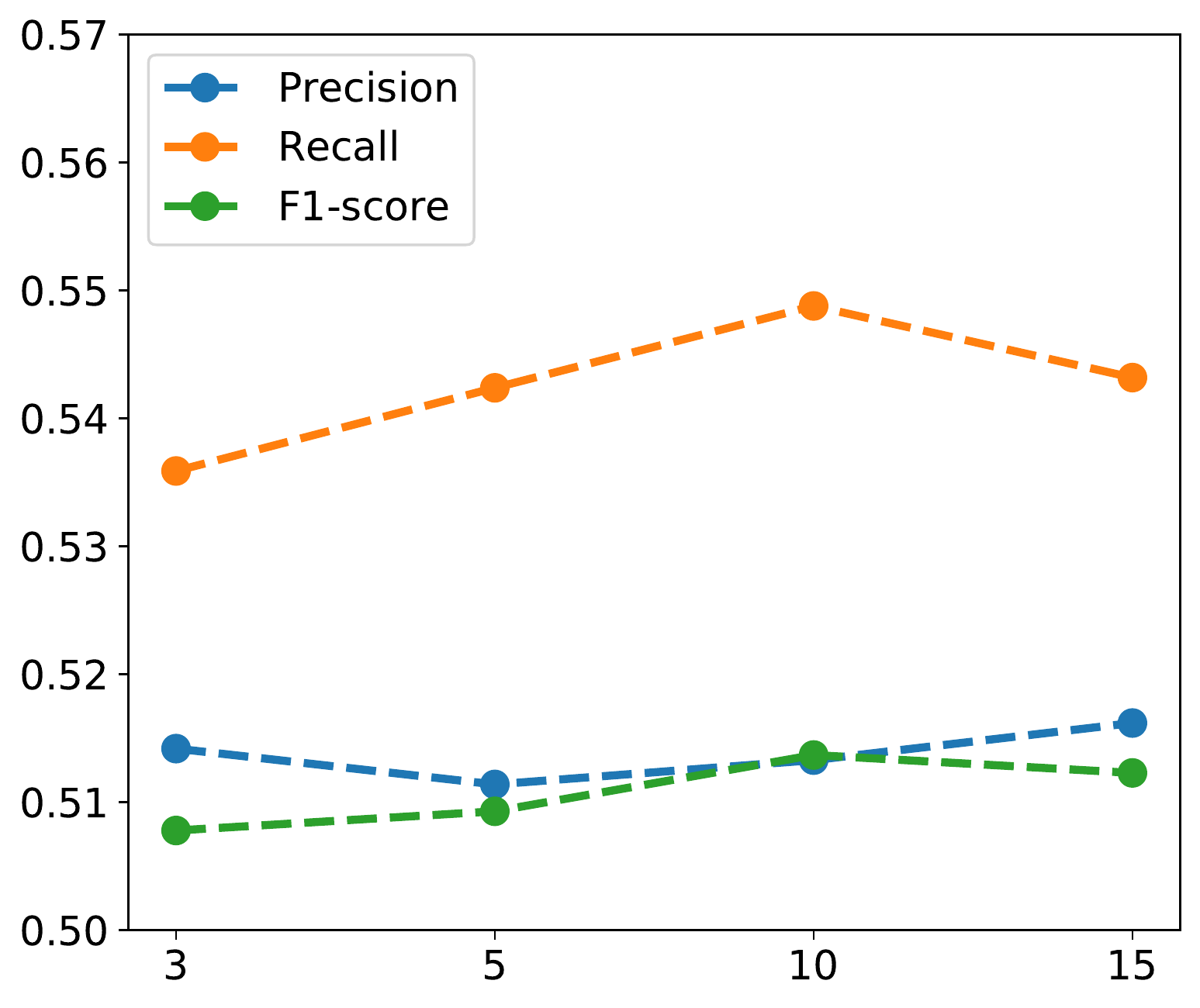}
    \label{fig:negative:1}
}
\subfigure[DPR-AG]{
    \includegraphics[scale=0.25]{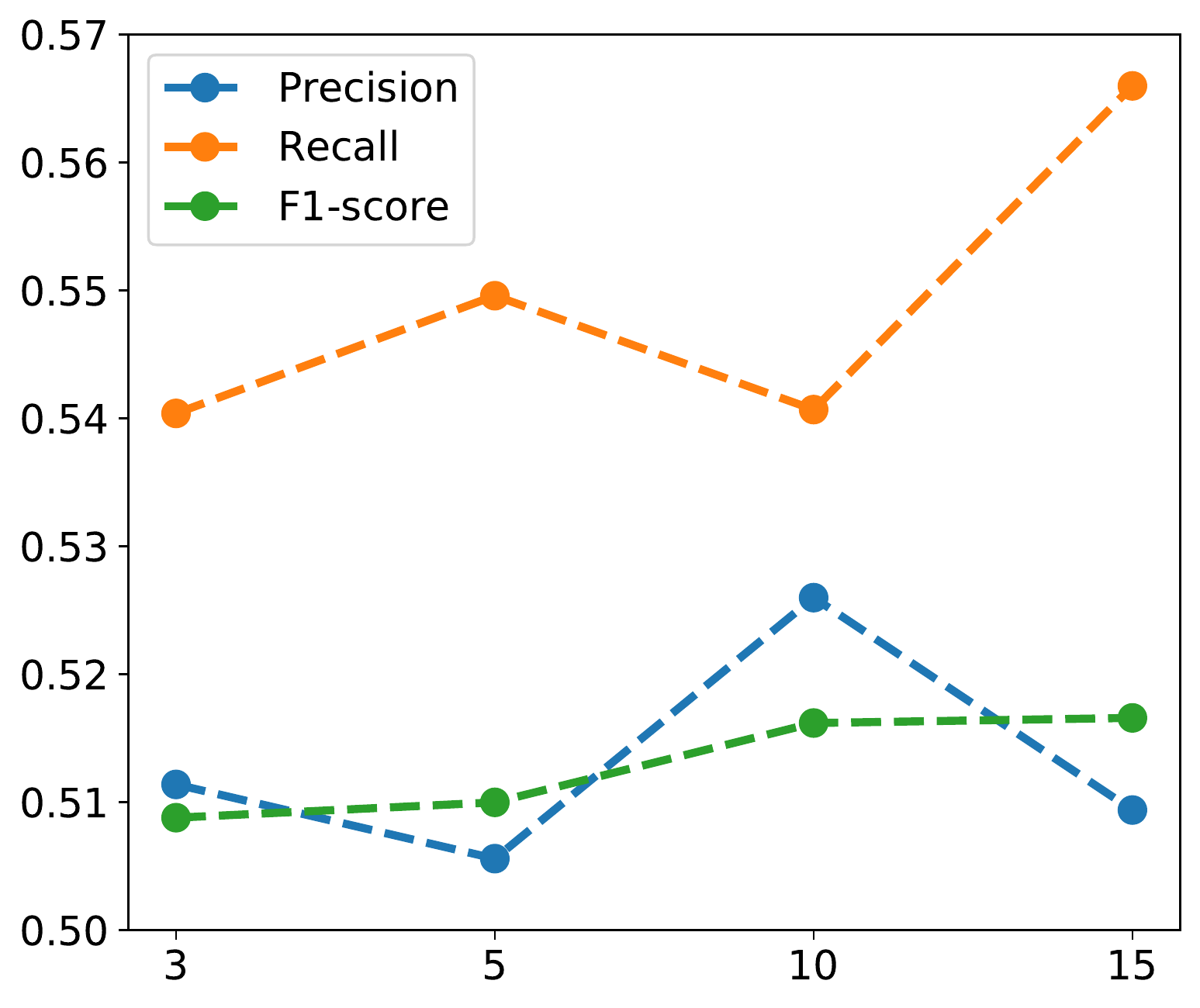}
    \label{fig:negative:2}
}
\caption{The performance of DPR-WG and DPR-AG with different number of negative samples.}
 \label{fig:negative}
\end{figure}

\subsection{Parameter Sensitivity}
We investigated the sensitivity of our model parameter in this section. 
First, we evaluated 
how the threshold for co-occurrence proportion affected the performance, and the results are 
shown in Figure~\ref{fig:threshold}. 
From the results, we can find that as the number 
of edges decreases, the model performance does not change significantly, and the F1-score 
shows a downward trend. This indicates the fact that there is no interaction 
between most of the drug pairs.

Next, we investigated whether utilizing two GNN layers can 
affect the results. Table~\ref{tab:layer} shows the results of our two models with 
one and two GNN layers. The results have not witnessed a performance improvement  
by adding one more GNN layer. As mentioned before, different from general graphs, we only need 
one GNN layer to extract almost all the information we expect since the drug package graph 
is dense enough.

Finally, we verified the impact of the negative sampling ratio. As shown in Figure~\ref{fig:negative}, 
we can find the performance only fluctuates in a small range, and the model with a 
small negative sample number also works well in practice. 
All the above experiments have proved 
the robustness of the methods proposed in this paper.

\subsection{Case Study}
\begin{figure}[tb]
  \centering
  \includegraphics[width=0.38\textwidth]{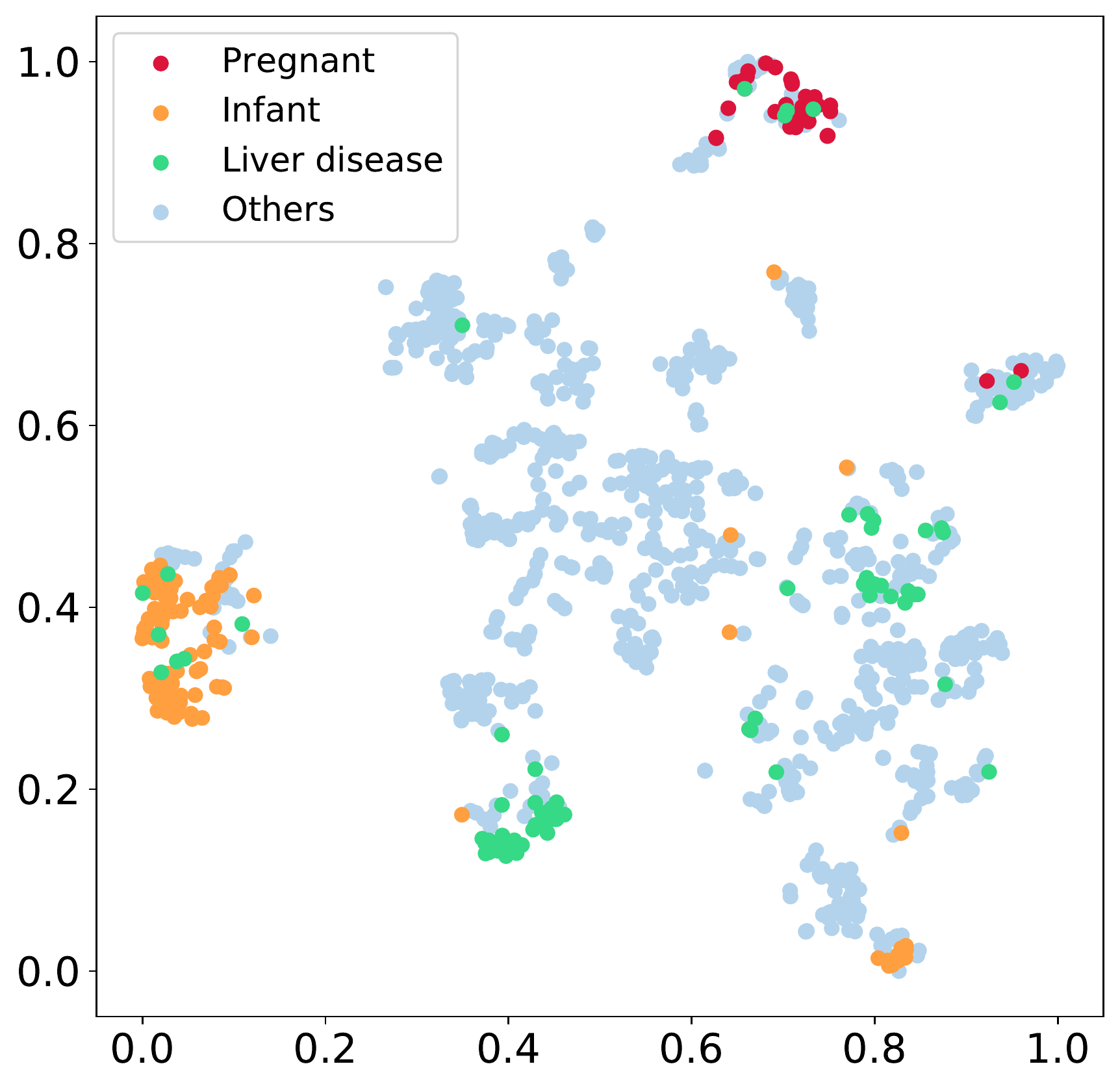}
  \caption{Visualization of mask vectors.}
  \label{fig:mask}
  \end{figure}
In this part, we present some cases to illustrate the effectiveness of 
our models and reveal some interesting medical rules based on the derived insights on 
patient conditions and drug interaction.

\subsubsection{Mask Vector Analysis}
As mentioned before, we extracted the mask vector 
$\sigma\left(MLP\left(\mathbf{u}\right)\right)$ of patient $u$ to describe the impact 
of the patient condition. In order to analyze the effect of the mask vectors, 
we randomly selected 1,000 patients and their corresponding mask vectors, and 
projected them into two-dimensional space with t-SNE, which is proposed in \cite{maaten2008visualizing}. 
We further selected three representative patient groups with special needs for drugs based on common 
sense, respectively pregnant women, infants (or young children) and patients with liver disease. 

Figure~\ref{fig:mask} shows the visualization result. 
We can find that 
the mask vectors of infants and pregnant women deviate the 
most from the vectors of other patients, which indicates that these two groups have 
the most special requirements for drug selecting, 
and this is consistent with our common sense. 
Moreover, the mask vectors of patients with liver disease are also relatively 
deviated from other patients, 
but the degree of aggregation is lower than previous two groups. This indicates 
that patients with liver disease have special needs for drugs, 
but there are also certain personalized needs. We can further study the impact of 
patient conditions on drug selection by statistical methods such as clusting, which 
shows a great possibility of our method to help medical researchers.

\subsubsection{Contextual Impact Factor Analysis} 
\begin{table*}[tb]
  \caption{Contextual Impact Factor Analysis for Patient \#27667.}
  \label{tab:caseWG}
  \begin{tabular}{ccccc}
  \toprule
  Drug 1&Drug 2&Description &Type&Factor\\
  \midrule
  Potassium Chloride& Cefazolin&drug 2 may decrease the excretion rate of drug 1.&Synergism&0.993\\
  Midazolam& Potassium Chloride&drug 1 may decrease the excretion rate of drug 2.&Synergism&-0.264\\
  Ephedrine&Methylprednisolone&drug 1 may increase the excretion rate of drug 2.&Antagonism&-0.309\\
 
  \bottomrule
  \end{tabular}
  \end{table*}

  \begin{table*}[tb]
    \caption{Edge Attribute Analysis for Patient \#25256.}
    \label{tab:caseAG}
    \begin{tabular}{cccccc}
    \toprule
    Drug 1&Drug 2&Type&$softmax\left(\mathbf{e}_{vu}^{\top}\mathbf{Q}\right)$&$softmax\left(\hat{\mathbf{e}}_{vu}^{\top}\mathbf{Q}\right)$\\
    \midrule
    Warfarin&Ondansetron&Synergism&[0.007, 0.807, 0.184]&[0.015, 0.923, 0.061]\\
    Metformin&Spironolactone&Antagonism&[0.358, 0.163, 0.478] &[0.769, 0.022, 0.208]\\
    
    \bottomrule
    \end{tabular}
    \end{table*}
In Section~\ref{sec:DPR-WG}, we propose to utilize contextual impact factors to 
reflect the impact of patient condition on drug interaction. 
In this section we will show 
how these impact factors play a role for recommending packages. 

We picked patient \#27667 for detailed analysis. From the EMR database we can know that 
this patient suffered from gallstones and came to the hospital for cholecystectomy. 
We input the ground truth drug package into DPR-WG, and got the contextual impact 
factors between all of the 
drug pairs. Table~\ref{tab:caseWG} shows three samples of drug pairs with different 
interaction types. From the results we can get several observations. 
First, Cefazolin can bring Potassium Chloride to a higher serum level, and the 
contextual impact factor for this edge is very high. This shows that our model believes 
the synergism between these two drugs is necessary for this patient. Second, 
Midazolam can also bring Potassium Chloride to a higher serum level, but our model 
gives a small negative factor for this interaction. By understanding more medical knowledge, 
we know that the combination of these two drugs has a greater risk, and they are even 
used for euthanasia. Our model predicts the risk of these two drugs. Finally, Ephedrine may decrease the effect of Methylprednisolone, 
but Methylprednisolone is an adrenal glucocorticoid with strong anti-inflammatory effect, 
which is very necessary for this patient. So our model gives a small negative factor to 
adjust the interaction effect between these two drugs. The above examples strongly 
confirm the effectiveness and interpretability of DPR-WG from different perspectives.

\subsubsection{Edge Attribute Analysis} 
In Section~\ref{sec:DPR-AG}, edge attribute vectors are calculated to describe the interaction 
between two drugs. The attribute vectors are forced to contain drug interaction category 
information, and mask vectors are utilized to bring the impact of patient condition. 
We propose that the mask vector plays a role by feature selecting. If we multiply a 
contextual edge attribute vector $\mathbf{e}_{vu}$ with the classification transfer matrix $\mathbf{Q}$, 
we can get a personalized drug interaction classification result, and we will illustrate 
this intuition in this case study.

We picked patient \#25256 for detailed analysis. This patient was a 79-year-old woman with 
high blood pressure, diabetes and heart disease. We got the corresponding patient mask 
vector and drug interaction vectors by DPR-AG, and we further got the 
non-personalized and personalized drug interaction classification results for the drug interaction 
vectors. Table~\ref{tab:caseAG} shows two examples for this. We can find that  
Warfarin and Ondansetron have a synergistic effect, and the initial drug interaction 
vector reflects this point. Furthermore, the mask vector enhances this feature, since 
Warfarin (which can prevent the formation and development of thrombus) is very 
important for this patient, and it is beneficial to keep this synergistic effect. 
In addition, Metformin and Spironolactone are marked as antagonistic, but it is not 
significantly reflected in the drug interaction vector, 
and the mask vector believes that 
these two drugs may have no interaction. To explain this, we consulted a doctor and 
learned that the interaction between these two drugs is relatively moderate, and 
they are often used together clinically. These examples clearly confirm the 
interpretability and learning ability of our model.

\section{PACKAGE GENERATION}
\begin{table}[tb]  
    \centering  
    \caption{The results of package generation.}  
    \label{tab:generation}  
      \begin{center}  
          \begin{tabular}{c|cc}  
              \hline  
            model & Non-heuristic & Heuristic \\\hline
            doctor1 & 39\% & 61\% \\
            doctor2 & 37\% &63\% \\
            doctor3 & 39\% &61\% \\
            doctor4 & 45\% &55\% \\
            doctor5 & 30\% &70\% \\\hline
            average & 38\% &62\%  \\
            
        \hline 
          \end{tabular}  
      \end{center}  
 
    \end{table}
\newcommand{\tabincell}[2]{\begin{tabular}{@{}#1@{}}#2\end{tabular}}  

\begin{table*}[tb]  
        \centering  
        \caption{Examples for the heuristic method.}
        \label{tab:heuristic}  
          \begin{center}  
              \begin{tabular}{c|ccc}  
                  \hline  
                Patient ID&Ground Truth&Non-heuristic & Heuristic\\\hline
                \#28062&\tabincell{c}{Glucose, Isoniazid,\\Silybin, Kanamycin,\\Rifampin, Levofloxacin,\\Aspirin, Clindamycin, Pyridoxine}&\tabincell{c}{Glucose, Levofloxacin\\Clindamycin, Moxifloxacin}&\tabincell{c}{Glucose, Levofloxacin\\Clindamycin, Moxifloxacin\\\textbf{Isoniazid}, \textbf{Silybin}, \textbf{Pyridoxine}}\\\hline
                \#28199&\tabincell{c}{Carboprost methyl ester,\\Cefuroxime Sodium,\\Aminomethylbenzene,\\Hydroxyethyl starch,\\Peptide hormones,Lidocaine}&\tabincell{c}{Lidocaine, Glucose,\\Cefuroxime, Aminomethylbenzene,\\ Carboprost methyl ester,\\Carprost tromethamine,\\Peptide hormones, Reserpine, Oxytocin}&\tabincell{c}{Carboprost methyl ester,\\Aminomethylbenzene, Misoprostol,\\\textbf{Hydroxyethyl starch},\\\textbf{Peptide hormones}, \textbf{Lidocaine}}\\
            \hline 
              \end{tabular}  
          \end{center}  
     
    \end{table*}
Until now, we have considered recommending drug packages 
that already exist within the EMR database. However, existing packages cannot 
meet the needs of new patients sometimes. Therefore, we present a heuristic algorithm which 
combines the existing packages, personalized drug prediction lists and drug interaction 
matrix to generate new packages. The algorithm is described as follows.

First, we get the drug frequency rank list $L$ which contains drugs in descending 
order of occurrence frequency in the EMR dataset. Then, we calculate 
the drug co-occurrence proportion matrix $M$ which is 
mentioned in Section~\ref{sec:package_graph}. 
For a new patient, we can get the patient embedding based on the 
patient's description. With the patient embedding, we can get the candidate set 
$S_1$ from similar patients as previously mentioned, and we can get the personalized 
prediction list $l$ of all drugs by utilizing the NCF model obtained in the pre-training 
phase, which contains drugs in descending 
order of predict value. It is worth noting that, as shown in Section~\ref{sec:overall}, 
the top drugs in $l$ can be incorrect. 
Finally, start with the initial candidate set $S_1$, 
we can get new drug packages as:

\begin{enumerate}
    \item Form a new candidate set $S_2$ based on reforming the packages in $S_1$ by the following ways:
    \begin{itemize}
        \item Delete the drugs that only appear in a small number of packages in $S_1$ and rank low in $l$;
        \item Add the drugs that rank low in $L$ and rank high in $l$, which means these drugs are not recommended just because they have high popularity.
    \end{itemize}
    \item Generate candidate set $S_3$ by modifying the drugs in $S_2$ using more radical strategies as:
    \begin{itemize}
        \item If drug $d$ ranks high in $l$ and has synergism relationship with a drug in package $p$, then add drug $d$ to package $p$;
        \item If drug $d$ ranks high in $l$ and has high co-occurrence proportion with a drug in package $p$, then add drug $d$ to package $p$;
        \item If drug $d_1$ and $d_2$ in package $p$ have antagonism relationship and low co-occurrence proportion, then delete the drug with lower lank in $l$;
    \end{itemize}
    \item returns final candidate set $S=S_1 \cup S_2 \cup S_3$.
\end{enumerate}

We verified the effectiveness of our heuristic algorithm on DPR-WG, where the non-heuristic model 
selected the best package from the initial candidate set $S_1$, and the heuristic model selected best package from $S$. 
Due to the hidden security risks of directly using the generated package, 
we randomly selected some test samples and handed them to five doctors to mark the 
packages they preferred. 

The results are shown in Table~\ref{tab:generation}, 
where the percentages reflect the ratio of the doctors' choice. From the 
results, we can find that utilizing the drug packages generated by the heuristic algorithm 
can significantly improve the performance of drug package recommendation. Furthermore, we picked 
two examples to illustrate the effect of the heuristic method. The examples are shown in 
Table~\ref{tab:heuristic}, where patient \#28062 is a patient with tuberculosis, and 
patient \#28199 is a pregnant woman. We can find that both the adding 
and deleting strategies are effective. For the adding strategy, Isoniazid was added since 
the rank promotion between list $L$ and $l$, and Pyridoxine was added because of the 
synergism interaction with Levofloxacin in the first example. 
For the deleting strategy, several incorrect drugs were deleted in the second example. 
All the results confirm the effectiveness of our package generation method.

\section{CONCLUSION}
In this paper, we studied the problem of drug package recommendation. Specifically, we 
first designed a pre-training method based on neural collaborative filtering to get 
the initial embedding of patients and drugs. Then, the drug interaction graph was 
initialized based on medical records and domain knowledge. Furthermore, we proposed a new 
drug package recommendation framework with two variants, respectively DPR-WG 
and DPR-AG to solve the problem, 
in which each the interactions was described as signed weights or 
attribute vectors. Finally, extensive experiments on a real-world 
data set from a first-rate hospital demonstrated the effectiveness of our DPR 
framework compared with several competitive baseline methods, and further supported 
the heuristic study for the drug package generation task with adequate performance.

\bibliographystyle{ACM-Reference-Format}
\bibliography{main}










\end{document}